\documentclass[a4]{article}
\usepackage{graphicx}
\usepackage{cite}
\usepackage{upgreek}
\usepackage{amsmath}
\usepackage{hyperref}

\usepackage{color}
\usepackage[normalem]{ulem}

\newcommand{\vM}{{\bf M}}

\newcommand{\vB}{{\bf B}}
\newcommand{\vA}{{\bf A}}

\newcommand{\vT}{{\bf T}}

\newcommand{\ve}{{\bf e}}

\newcommand{\vr}{{\bf r}}

\newcommand{\degr}{^{\circ}}

\begin{document}

\title{3D modeling of a Superconducting Dynamo-Type Flux Pump}

\author{Asef~Ghabeli$^1$,Enric~Pardo$^1$\footnote{Corresponding author: enric.pardo@savba.sk},Milan~Kapolka$^{1,2}$\\
\normalsize{$^1$Institute of Electrical Engineering, Slovak Academy of Sciences,}\\
\normalsize{Bratislava, Slovakia}\\
\normalsize{$^2$University of Leicester, Aerospace $\&$ Computational Engineering,}\\
\normalsize{University Road, LE1 7RH, Leicester, UK}
}

\maketitle


\begin{abstract}
High temperature superconducting (HTS) dynamos are promising devices that can inject large DC currents into the winding of superconducting machines or magnets in a contactless way. Thanks to this, troublesome brushes in HTS machines or bulky currents leads with high thermal losses will be no longer required. The working mechanism of HTS dynamo has been controversial during the recent years and several explanations and models have been proposed to elucidate its performance. In this paper, we present the first three-dimensional (3D) model of an HTS flux pump, which has good agreement with experiments. This model can be beneficial to clarify the mechanism of the dynamo and pinpoint its unnoticed characteristics. Employing this model, we delved into the screening current and electric field distribution across the tape surface in several crucial time steps. This is important, since the overcritical screening current has been shown to be the reason for flux pumping. In addition, we analyzed the impact of both components of electric field and screening current on voltage generation, which was not possible in previous 2D modeling. We also explored the necessary distance of voltage tab at different airgaps for precise measurement of the voltage across the tape in the dynamo.
\end{abstract}

\flushbottom
%
%
\thispagestyle{empty}


\section{Introduction}
       	
Since 1960s, when the first generation of superconducting flux pumps have been designed and operated, the flux pump technology has undergone a remarkable developement \cite{van1981fully,britton1972fluxpumps,coombs2019superconducting,coombs2016overview,pardo2017dynamic}. Within around 60 years, various mechanically or electrically driven superconducting flux pumps have been operated using low-temperature superconductor (LTS) materials in their previous generation and high-temperature superconductor (HTS) materials in their last generation. 

Superconducting dynamos convert mechanical energy into electromagnetic energy. They use permanent magnets as the source of varying magnetic field to induce DC volatge inside a superconducting wire. In their previous generation with LTS wires, creating normal region in superconductor was necessary to generate voltage and the local magnetic field needed to exceed the critical magnetic field of the wire. Generating this high amount of magnetic field in commercial LTS wires is not always possible, which limits the application of LTS superconducting dynamos. However, since Hoffmann and \textit{et al.} designed and implemented the new generation of flux pumps with HTS materials \cite{hoffmann2010flux}, the application of flux pumps and in particular HTS dynamos has became popular and widespread. In this type of dynamos, the existence of normal regions in superconductor is no more required and the voltage can be created even with very low amount of alternating magnetic field \cite{Ghabeli_2020}. The only necessary characteristic for a material to be used in the new generation of flux pump is having non-linear resistivity \cite{mataira2019origin}. This makes the HTS dynamos a reliable and efficient candidate to be utilized in superconducting electrical machines \cite{bumby2016development,kalsi2006status,badcock2016impact,pantoja2016impact,
hamilton2018design,storey2019optimizing,jiang2015impact,bumby2016through}.          

The simplified working mechanism of an HTS flux pump can be explained as follows. Any magnitude of varying magnetic field causes circulating screening currents in the tape surface. These screening currents result in electric field in the areas above the local critical current of the tape. The accumulation of this induced electric field in one complete cycle results in certain net voltage, which causes flux pumping. However, this is not the case in materials with linear resistivity, where the net voltage is always zero. In other words, having non-linear resistivity is the key and the only necessary characteristic of a material to be used in an HTS flux pump. In addition, the local critical current is subjected to alternation based on the local magnetic field density and its angle with respect to tape surface. The $J_c(B,\theta)$ dependency can be measured via experiment, where $\theta$ is the magnetic field angle with respect to the tape surface.           
 
Although the mechanism of HTS dynamo have been investigated and explained in \cite{mataira2019origin,Ghabeli_2020,matairamechanism} using experiments and 2D modeling, there are still many aspects that have been left untouched. These issues can hardly be disclosed by experiments and are not feasible to be studied via 2D modeling. The first one is the complete distribution of screening current on tape surface during flux pumping process. The current density and electric field on the tape surface have two components: $J_x$, $E_x$ along the tape width and $J_y$, $E_y$ along the tape length. 2D modeling assumes infinitely long tape along the length, thus there are only $J_y$ and $E_y$ components for current density and electric field and $J_x$ and $E_x$ are missing. Considering these facts and limitations of 2D modeling, the distribution of overcritical screening current, which has been shown to be the reason for flux pumping and its resultant electric field was not thoroughly pinpointed in previous literature. That is why the necessity of a 3D model to certify the previous proposed mechanisms of flux pumping and to better clarify the procedure is inevitable. By using an accurate and efficient 3D model, it is possible to pinpoint the mechanism of flux pumping and its details. In addition, 3D modeling gives the opportunity to analyze more complex geometries, regarding both tape or magnet shapes, which is also not possible via 2D modeling.  
	
	In previous literature, there have been several works focusing on modeling of flux pumps in order to investigate its performance. In \cite{campbell2017finite} a simplified 2D model using $A$-formulation was proposed, which could simulate the HTS dynamo mechanism. Although this model could describe some key aspects of flux pumping, it was unrealistic. In \cite{mataira2019origin} a 2D finite element numerical model based on H-formulation was presented, which had fairly good agreement with experiment and could explain well the mechanism behind the voltage generation in an HTS dynamo in open-circuit condition. An efficient and fast model, yet with good agreement with experiment based on minimum electromagnetic entropy production method (MEMEP) was proposed to model the procedure of voltage generation in an HTS dynamo in open-circuit mode \cite{Ghabeli_2020}. In \cite{matairamechanism}, Mataira and \textit{et al.} extended their 2D model based on H-formulation to current-driven mode to further explain the underlying physics of voltage generation in an HTS dynamo. They also validated their modeling results with experiments. In \cite{ainslie2020new} a benchmark of several different models for modeling of HTS dynamo in open-circuit mode was presented. This benchmark includes $H$-formulation with shell current method \cite{mataira2019origin,mataira2020modeling,matairamechanism}, MEMEP method \cite{Ghabeli_2020}, coupled $H$-$A$ method \cite{brambilla2018finite}, coupled $T$-$A$ 
formulations \cite{benkel2020t}, Segregated $H$-formulation \cite{queval2018superconducting}, integral equation \cite{brambilla2009ac} and
volume integral equation-based equivalent circuit method \cite{morandi2014unified}. All of the methods showed very good agreement with each other either qualitatively or quantitatively and the MEMEP method was the fastest in terms of computational speed.	    

In this article, we present the first 3D model of a dynamo-type HTS flux pump based on MEMEP method. This method utilizes optimum number of degrees of freedom by solving all the variables of the problem only inside the superconducting region, which makes the model very fast and efficient. Employing the proposed model, the performance of the dynamo is investigated and analyzed. The article structure is as follows. First, we explain the modeling method, including the introduction of the MEMEP 3D method, magnet modeling process and the model configuration. Afterwards, the article analyses the screening current and electric field in several critical time steps of the magnet movement over the HTS tape. Then, the impact of length on voltage generation and measurement is investigated and, at the end, the model results are compared to experiments in various airgaps.     

\section{Dynamo Modeling}

\subsection{MEMEP 3D method} 
\label{MEMEP 3D method}

In this article, we model the HTS dynamo by means of the Minimum Electro-Magnetic Entropy Production method in 3D (MEMEP 3D). This variational method is based on $\textbf{T}$-formulation and is able to exploit several strategies to speed up calculation such as parallel computing, dividing into sectors and symmetry. The method works based on minimizing a functional containing all the variables of the problem including magnetic vector potential $\textbf{A}$, current density $\textbf{J}$ and scalar potential $\varphi$. It is proven that the minimum of the 3D functional for any given time step is the unique solution of Maxwell differential equations \cite{Pardo16IES,pardoE2017JCP,Kapolka18IESa,kapolka20193d}. 

For magnetization problems, the method uses the effective magnetization $\textbf{T}$ as state variable, defined as
\begin{equation}
\nabla\times \textbf{T}=\textbf{J}, 
\end{equation}
where $\textbf{J}$ is the current density. Configurations with transport current are possible after adding an additional term to the equation above \cite{pardoE2017JCP,milan_kapolka_2019_4003453}. The general relation between the electric field and vector and scalar potentials is
\begin{equation}
\textbf{E}(\textbf{J})=-\partial_t{\textbf{A}}-\nabla\varphi, 
\label{E(J)}
\end{equation} 
and the current conservation equation is
\begin{equation}
\nabla\cdot\textbf{J}=0, 
\label{nablaJ}
\end{equation}
where $\partial_t{\textbf{A}}$ is the change of vector potential with respect to time at a given location $\vr$ and $\varphi$ is the scalar potential. Equation (\ref{nablaJ}) is always satisfied because $\nabla\,\cdot\,(\nabla\times\textbf{T})=0$. Thus, we only need to solve equation (\ref{E(J)}). In addition, the Coulomb's gauge, $\nabla\cdot\textbf{A}=0$, has been assumed to solve the 3D problem, and hence $\varphi$ becomes the electrostatic potential \cite{Grilli2014computation}. 

Since the current density $\textbf{J}$ and $\vT$ only exist inside the material, meshing is only needed in this region, increasing the calculation speed significantly \cite{pardoE2017JCP}. The vector potential $\textbf{A}$ in the functional has two contributions including $\textbf{A}_{a}$ and $\textbf{A}_{J}$ standing for the vector potential due to the applied field and the vector potential due to the current density in the superconductor, respectively. The ${\bf A}_a$ component can be replaced by the vector potential due to the magnet in a dynamo, while $\textbf{A}_{J}$ is calculated with the following volume integral of current density 
\begin{equation}
\textbf{A}_{J}(\textbf{r})=\frac{{\mu}_{0}}{4\pi} \int_{V}dV' \dfrac{\textbf{J}(\textbf{r}')}{|\textbf{r}-\textbf{r}'|}=\frac{{\mu}_{0}}{4\pi} \int_{V}dV' \dfrac{{\nabla}'\times \textbf{T}(\textbf{r}')}{|\textbf{r}-\textbf{r}'|}. 
\label{eq2.1}
\end{equation}

As it was demonstrated in \cite{pardoE2017JCP}, solving equation (\ref{E(J)}) is equivalent to minimizing the following functional
\begin{eqnarray}
L=\int_{V}dv[\frac{1}{2} \frac{\Delta \textbf{A}_{J}}{\Delta t}\cdot(\nabla \times \Delta \textbf{T})+ \frac{\Delta \textbf{A}_{a}}{\Delta t}\cdot (\nabla \times \Delta \textbf{T})+U(\nabla \times \textbf{T})],
\label{functional}
\end{eqnarray}
where $U$ is dissipation factor, which can include any $\textbf{E}-\textbf{J}$ relation of superconductor
\begin{equation}
U(\textbf{J})=\int_{0}^{J}\textbf{E}(\textbf{J}')\cdot\textbf{J}'.
\label{dissipation_factor}
\end{equation}

To solve the problem in time dependent mode, the functional is minimized in discrete time steps. Assuming the functional variables in a certain time step $t_{0}$ as $\textbf{T}_{0}$, $\textbf{A}_{J0} $ and $\textbf{A}_{a0}$, the upcoming time step will be $t=t_{0}+\Delta t$ and the variables in the time step $t$ will be $\textbf{T}=\textbf{T}_{0} +\Delta\textbf{T}$, $\textbf{A}_{J}=\textbf{A}_{J_0}+ \Delta \textbf{A}_{J}$ and $\textbf{A}_{a}=\textbf{A}_{a0}+ \Delta \textbf{A}_{a}$, respectively, where $\Delta \textbf{T}$, $\Delta \textbf{A}_{T}$ and $ \Delta \textbf{A}_{a}$ are the variables difference between two time steps and $\Delta t$ is the time difference between the two time steps, which does not need to be uniform.

The optimum number of mesh for a tape with dimensions 12$\times$48$\times$0.001 in mm is chosen as 50$\times$60$\times$1 along the $x$, $y$ and $z$ axis, respectively inside the superconductor. The results remain almost unchanged for higher number of meshes. However, for having better resolution on the tape surface, instead of 60 mesh, 200 meshes has been chosen along the tape length. The thin film approximation (one mesh along the tape thickness) has been considered since it has no effect on the results due to low thickness of the tape. However, the model enables to take several elements in the thickness into account \cite{kapolkaM2020SST}. 


\subsection{3D Cylindrical Magnet Modeling} 

	As explained in section \ref{MEMEP 3D method}, for our MEMEP 3D method, we need to calculate the vector potential due to the magnet as $\textbf{A}_a$ in the functional. Therefore, all we need is to calculate the vector potential and magnetic field of the magnet in certain observation points inside the tape, which are the central locations of each cell.

\begin{figure}[tbp]
\centering
{\includegraphics[trim=0 0 0 0,clip, width=8 cm]{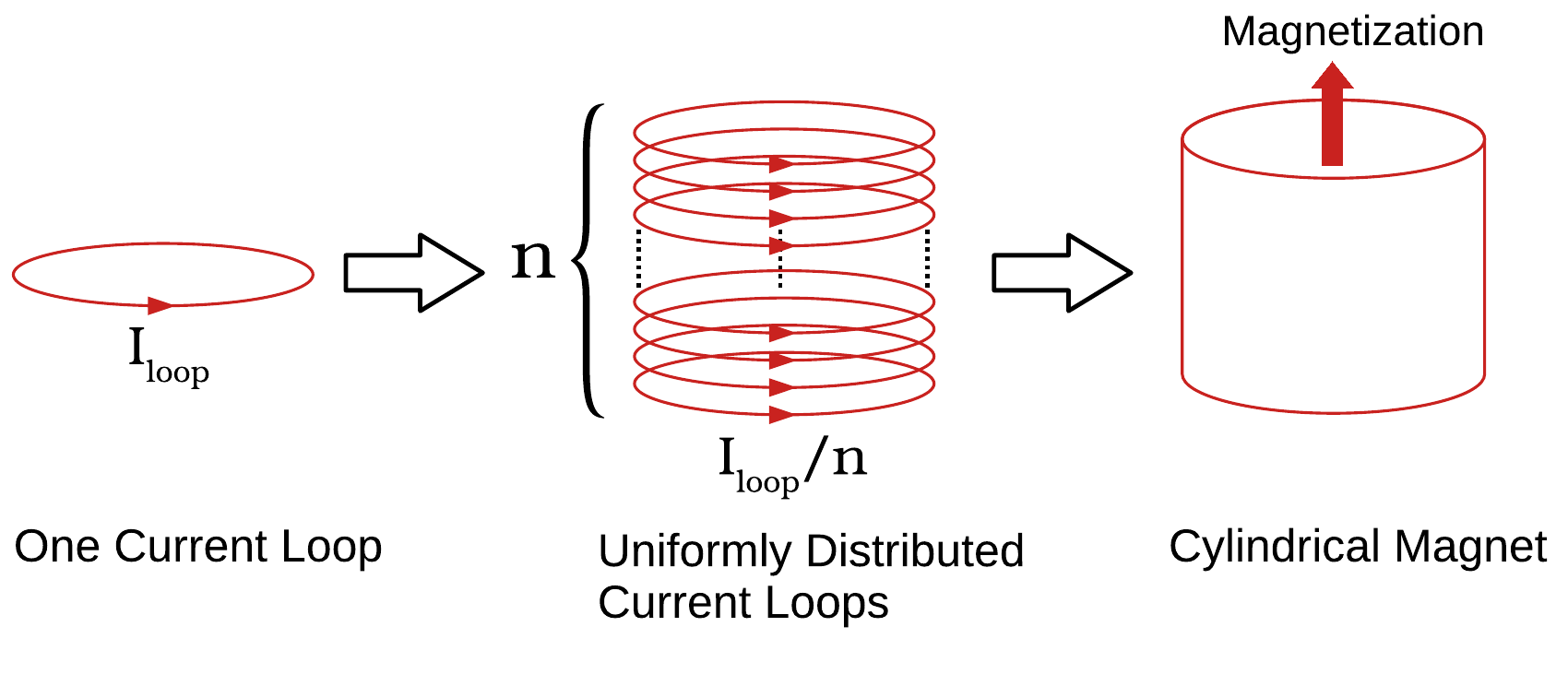}}
\caption {Simplified process of modeling a 3D cylindrical magnet by superposition of uniformly distributed current carrying loops.} 
\label{Magnet_Loops}
\end{figure}

The magnetic flux density, $\vB$, and vector potential, $\vA$, generated by a uniformly magnetized body is the same as those generated by the effective surface magnetization current density, ${\bf K}_M$, as
\begin{equation}
\textbf{K}_M=\vM\times\textbf{e}_n ,
\end{equation}  
where $\vM$ is the magnetization and $\ve_n$ is the unit vector perpendicular to the surface pointing outwards. For the cylindrical magnet, $\ve_n$ follows the $z$ direction in the cylindrical coordinate system with origin at the magnet center. Then, ${\bf K}_M$ follows the angular direction, ${\bf K}_M=K_M\ve_\phi$. We numerically calculate $\vB$ and $\vA$ from the magnet by superposition of many current loops distributed uniformly along the magnet height \cite{simpson2001simple,domokoscomputation} with current $I_l=K_M\Delta z$, where $\Delta z$ is the separation between loops. The number of current loops will be increased until it reaches to a point that with further increasing, the value of vector potential in observation points remain unchanged. This process leads to a high accuracy for calculation of magnet vector potential in observation points. Fig. \ref{Magnet_Loops} illustrates this process. 

The vector potential of the magnet, $\vA$, is calculated by the sum of all the small loops. Since the magnet could have any orientation and position with respect to the general Cartesian axis, $\vA$ at any location is obtained by rotation and translation. As a result, $\vA$ may have all 3 Cartesian components in a general reference axis \cite{landau2013electrodynamics,callaghan1960magnetic}. However, this vector potential has only angular component, $A_\phi$, regarding cylindrical coordinates centered on the magnet $(r,z,\phi)$, being  
\begin{equation}
\label{Aphi}
A_\phi=\frac{\mu_0 I_l}{4\pi}\sqrt{\frac{a}{r}}\left[\left(1-\frac{1}{2}k^2\right)K(k)-E(k)\right] ,
\end{equation} 
where $\mu_0$ is the permeability of vaccum, $I_l$ is the loop current, $a$ is the current loop radius, $k^2=4ar/[(a+r)^2+z^2]$ and $K(k)$ and $E(k)$ are complete elliptic integrals of the first and second kinds, respectively \cite{galassi2002gnu}. Converting $A_{\phi}$ from cylindrical coordinate system to Cartesian coordinate system gives rise to two components of $A_x$ and $A_y$ for the current loop. However, by applying rotation to the magnet in Cartesian coordinate system, another component will also be arisen as $A_z$.  

For calculations regarding $J_c(B,\theta)$, the components of magnetic field due to the magnet in the tape are also required, which can be obtained with the same process of superposition of current carrying loops as vector potential. The magnetic field due to a current loop in its local cylindrical coordinate system has only two components as $B_r$ and $B_z$, as follows, while $B_\phi=0$ \cite{landau2013electrodynamics,domokoscomputation,callaghan1960magnetic}
\begin{equation}
\label{Br}
B_r=\frac{\mu_0 I_l}{4\pi}\frac{2z}{r\sqrt{[(a+r)^2+z^2]}}\left[-K(k)+\frac{a^2+r^2+z^2}{(a-r)^2+z^2}E(k)\right]
\end{equation}
\begin{equation}
\label{Bz}
B_z=\frac{\mu_0 I_l}{4\pi}\frac{2}{\sqrt{[(a+r)^2+z^2]}}\left[K(k)+\frac{a^2-r^2-z^2}{(a-r)^2+z^2}E(k)\right] .
\end{equation} 

The modeled permanent magnet selected for this study is a cylinderical magnet with the dimension of 10$\times$10$\times$10 mm \cite{bumby2016anomalous}. The magnet type is a N42 (Nd-Fe-B magnet), possessing the remanence magnetic field of around 1.3 T.  


\subsection{HTS Tape modeling}
\label{HTS Tape modeling}
	The isotropic $\textbf{E}-\textbf{J}$ power law is implemented in the functional within the dissipation factor of (\ref{dissipation_factor}), assuming the following non-linear characteristic of the superconductor	
\begin{equation}
\textbf{E}(\textbf{J})=E_{c}\left(\frac{|\textbf{J}|}{J_{c}}\right)^n \frac{\textbf{J}}{|\textbf{J}|} ,
\label{eq5}
\end{equation}
where $E_{c} =10^{-4}$ V/m is the critical electric field, $J_{c}$ is the critical current density and $n$ is the n-value of the HTS tape. 
                        
The parameters of the modeled Superpower SF12050CF wire including $J_{c}(B,\theta)$ data are derived from \cite{mataira2019origin}. The tape has 12 mm width, 1 $\mu$m thickness, 48 mm length, n-value of 20, and the critical current at self field of 281 A. The modeled tape has similar (but not identical) characteristics to the HTS tape used in \cite{bumby2016anomalous}, where we aim to compare our modeling results with the experimental results of that article. 

\begin{figure}[tbp]
\centering
{\includegraphics[trim=0 0 0 0,clip,width=12 cm]{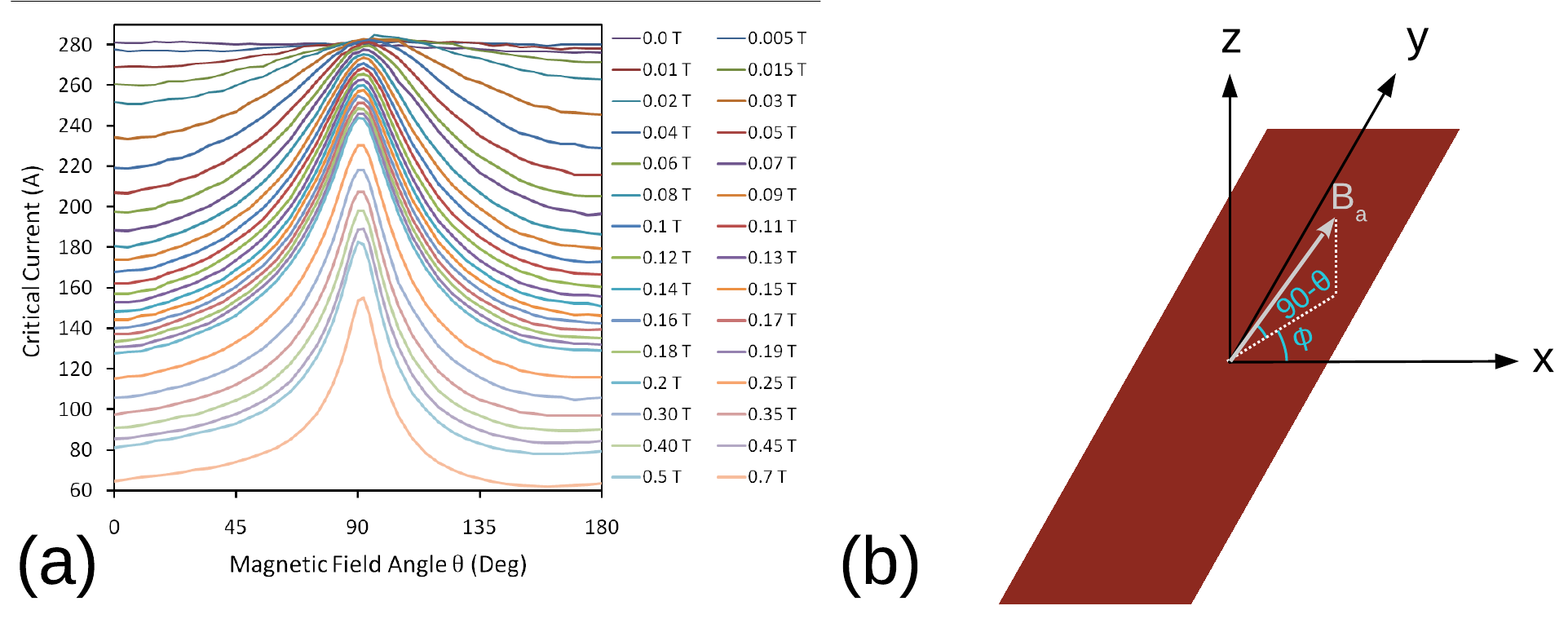}}
\caption { (a) Experimental $I_{c}(B,\theta)$ data used as input $J_c$ for modeling, derived from \cite{mataira2019origin}. Data was measured at 77.5 K in magnetic fields up to 0.7 T from a Superpower SF12050CF wire. (b) Sketch of $I_{c}(B,\theta,\phi)$ showing the angle $\theta$ along with angle $\phi$ as the angle between the $x$ axis and the projection of the applied magnetic field to the $xy$ plane. } 
\label{fig.angles}
\end{figure}

Fig. \ref{fig.angles}(a) shows the experimental $J_{c}({B},\theta)$ dependence, where $\theta$ is the angle of the applied magnetic field with respect to the normal vector of the tape surface [Fig. \ref{fig.angles}(b)]. Since the measurements were not performed in the whole 360$\degr$ range, we only use the measured critical current in the range between 0$^{\circ}$ to 180$^{\circ}$ and then we assume that is is symmetrical in the range between -180$^{\circ}$ to 0$^{\circ}$ for the model. In contrast to 2D modeling, in 3D modeling there is another angle, showed as $\phi$ in Fig. \ref{fig.angles}(b), which is the angle of applied magnetic field with respect to the $x$ axis. Although our modeling method allows this kind of $J_{c}(B,\theta,\phi)$ dependence \cite{kapolka20193d}, the $J_c$ measurements were done with $\phi=\pi/2$ only. Indeed, complete $J_{c}(B,\theta,\phi)$ measurements are scarce for any type of sample due to experimental complexity. Therefore, our model only considers the angle $\theta$, while angle $\phi$ is neglected with the assumption that $B_x$ plays the same role as $B_y$. In other words, we assume that $J_c(B,\theta,\phi)=J_c(B,\theta,\pi/2)$ for any $\phi$. As seen for other types of experiments, such as cross-field demagnetization, this assumption does not have a severe influence on the electromagnetic behavior \cite{kapolkaM2020SST}.

\subsection{Model Configuration}

\begin{figure}[tbp]
\centering
{\includegraphics[trim=0 0 0 0,clip,width=12.5 cm]{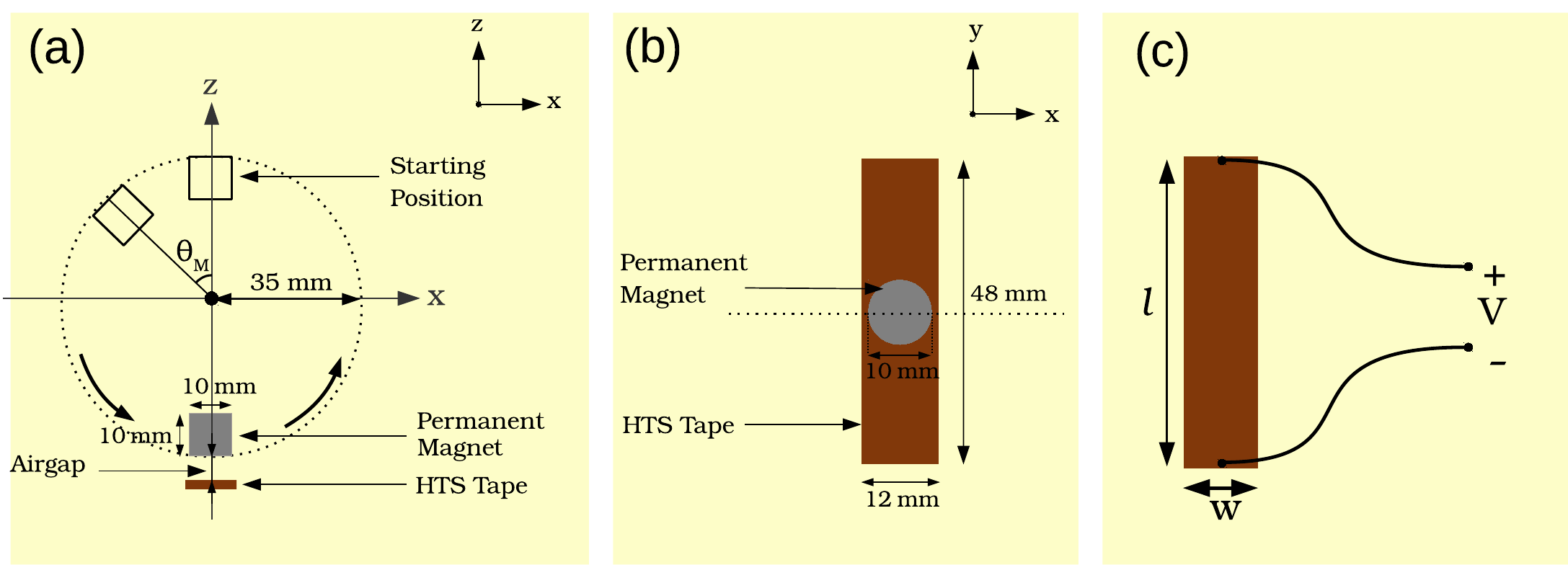}}
\caption {Configuration of the 3D model: (a) view from $xz$-plane, (b) view from $xy$-plane, (c) qualitative sketch of the voltage taps.} 
\label{fig.configuration}
\end{figure}

Fig. \ref{fig.configuration} shows the configuration of the 3D model in 2 different views. The magnet rotates in the $xz$-plane on top of the tape in counter clockwise direction, where the rotor external radius is 35 mm. The magnet magnetization is pointed out to the outside of rotation circle (toward the tape). $\theta_M$ is defined as the magnet angle, where the magnet starting position ($\theta_M=0$) is set when its center is aligned with the positive direction of $z$-axis. The airgap is defined as the minimum distance between the magnet outer surface and tape upper surface happening at ($\theta_M=180$). The rotating frequency is set to 12.3 Hz to be comparable with measurements conducted in \cite{bumby2016anomalous}. 

The calculation time for this configuration taking the $J_c(B,\theta)$ dependency into account and meshes 50$\times$200$\times$1 and 50$\times$60$\times$1 is around 16 hours and 40 minutes and less than 3 hours, respectively, on a desktop computer with two Intel(R) Xeon(R) CPUs of E5-2630 v4 $@$2.20GHz with 20 virtual cores (10 physical cores) each and 64 GB RAM. For the case of constant $J_c$ with 50$\times$200$\times$1 mesh the computation reduces to around 6 hours and 40 minutes.     

\subsection{General Definitions} \label{General Definitions}

In this article, we take into account that the voltage taps are placed at the ends of the, they are made of a conductor of negligible resistance, and they are very far away from the magnet at any time [Fig. \ref{fig.configuration}(c)]. With these assumptions, the output instantaneous open-circuit voltage of the flux pump is calculated using equation (\ref{E(J)}) as
\begin{eqnarray}
V=-\dfrac{1}{wl}\left( \int_{-\frac{w}{2}}^{+\frac{w}{2}}dx 
\int_{-\frac{l}{2}}^{+\frac{l}{2}}dy \left[\frac{\partial \varphi}{\partial y}\right]\right).\;l=\dfrac{1}{w}\int_{-\frac{w}{2}}^{+\frac{w}{2}}dx 
\int_{-\frac{l}{2}}^{+\frac{l}{2}}dy \left[\partial_t{\textbf{A}_y}+\textbf{E}_y(\textbf{J})\right],
\end{eqnarray}                   
where $l$ is the tape length and $w$ is the tape width, and $\textbf{A}_y$ and $\textbf{E}_y$ are $y$ components of total vector potential and electric field, respectively. The DC value of the instantaneous open-circuit voltage is
\begin{equation}
\label{DC value}
V_{DC}=f \int_{0}^{1/f} V(t)\; dt 
\end{equation} 
where $f$ is the rotation frequency of the magnet. $V_{DC}$ only depends on the electric field generated by the non-linear resistivity of tape, since the total vector potential $\textbf{A}$ is periodic in a full cycle \cite{Ghabeli_2020}. 

Another important parameter is $\Delta V$ since it can be obtained directly from measurements, which is defined as output voltage difference of the tape at 77$^{\circ}$ K (superconducting state) and at 300$^{\circ}$ K (normal state) 
\begin{eqnarray}
\label{DeltaV}
\Delta V (t)=V_{77^{\circ} K}(t)-V_{300^{\circ} K}(t) .
\end{eqnarray}
In other words, $\Delta V$ eliminates the contribution of induced electric field due to the vector potential from the magnet and considers only the contribution from the non-linear resistivity of the superconducting tape, since the eddy currents in the tape at normal state are negligible (at least, when the tape is not copper stabilized). In open-circuit mode, $\Delta V$ follows \cite{Ghabeli_2020}
\begin{equation}
\label{deltav}
\Delta V_{oc} \approx [\partial_t A_{av,J}+E_{av}(J)]. \;l
\end{equation}
where $V_{oc}$ denotes the open-circuit voltage, $A_{av,J}$ is the volume-average vector potential due to superconducting screening current, and $l$ is the tape length. 

\section{Modeling Results and Discussion}

\subsection{Current and Electric Field Analysis}
\label{Current and Electric Field Analysis}

In this section, all the modeling results have been calculated and presented using constant $J_c$, since it eliminates the complications of $J_c$ variations due to inhomogeneous magnetic field from the magnet and simplifies the performance study of the dynamo. Later, in section \ref{s.comp_exp}, we take the $J_c(B,\theta)$ dependence into account.
     
\begin{figure}[tbp]
\centering
{\includegraphics[trim=0 0 0 0,clip,width=10 cm]{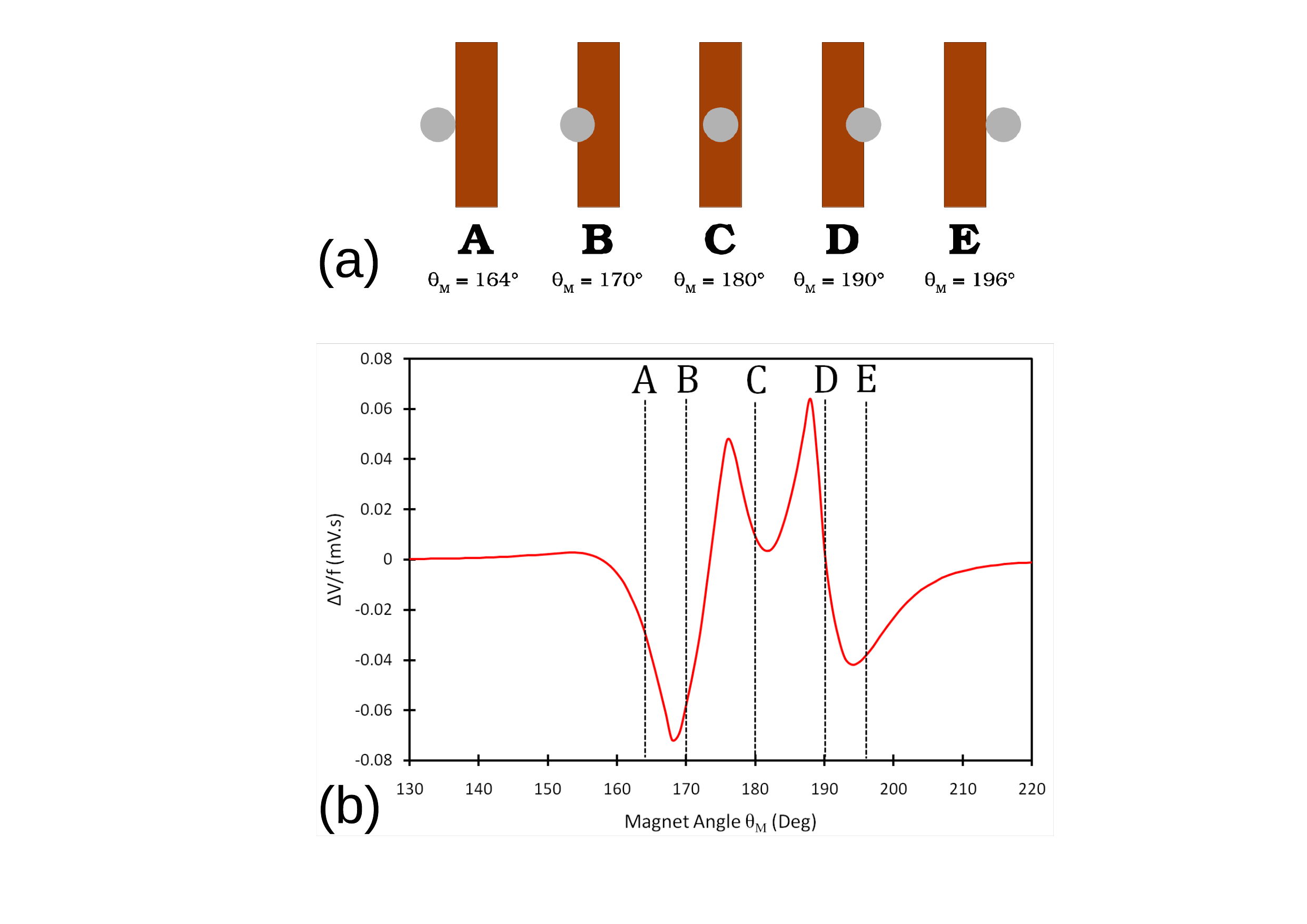}}
\caption {(a) Sketch of key magnet positions along with their magnet rotation angles $\theta_{M}$ from Fig. \ref{fig.configuration}. (b) Frequency-normalized $\Delta V$ versus magnet angle $\theta_{M}$ for airgap of 3.3 mm. } 
\label{fig.DeltaV_key}
\end{figure}

To examine better the flux pump behavior, we defined several key positions of the magnet while traversing the HTS tape. Fig.\ref{fig.DeltaV_key}(a) shows the schematic of these positions along with their magnet rotation angle $\theta_{M}$. These positions belong to the second cycle of the magnet rotation when the tape is already fully saturated with screening current and has passed the transient state. Fig. \ref{fig.DeltaV_key}(b) shows the graph of frequency-normalized $\Delta V$ versus magnet angle $\theta_{M}$. As it can be seen, in position \textbf{A}, while the magnet touches the tape, pumping voltage has already started. This is because the magnetic field of the magnet already appeared inside the tape before the magnet touches the tape, starting from around $\theta_{M}$=154$^{\circ}$. At position \textbf{C}, although the magnet is in the middle of the tape, the curve minimum happens shortly afterwards, which can be explained by the hysteresis nature of the tape screening currents.    

\begin{figure}[tbp]
\centering
{\includegraphics[trim=0 0 0 0,clip,width=12.5 cm]{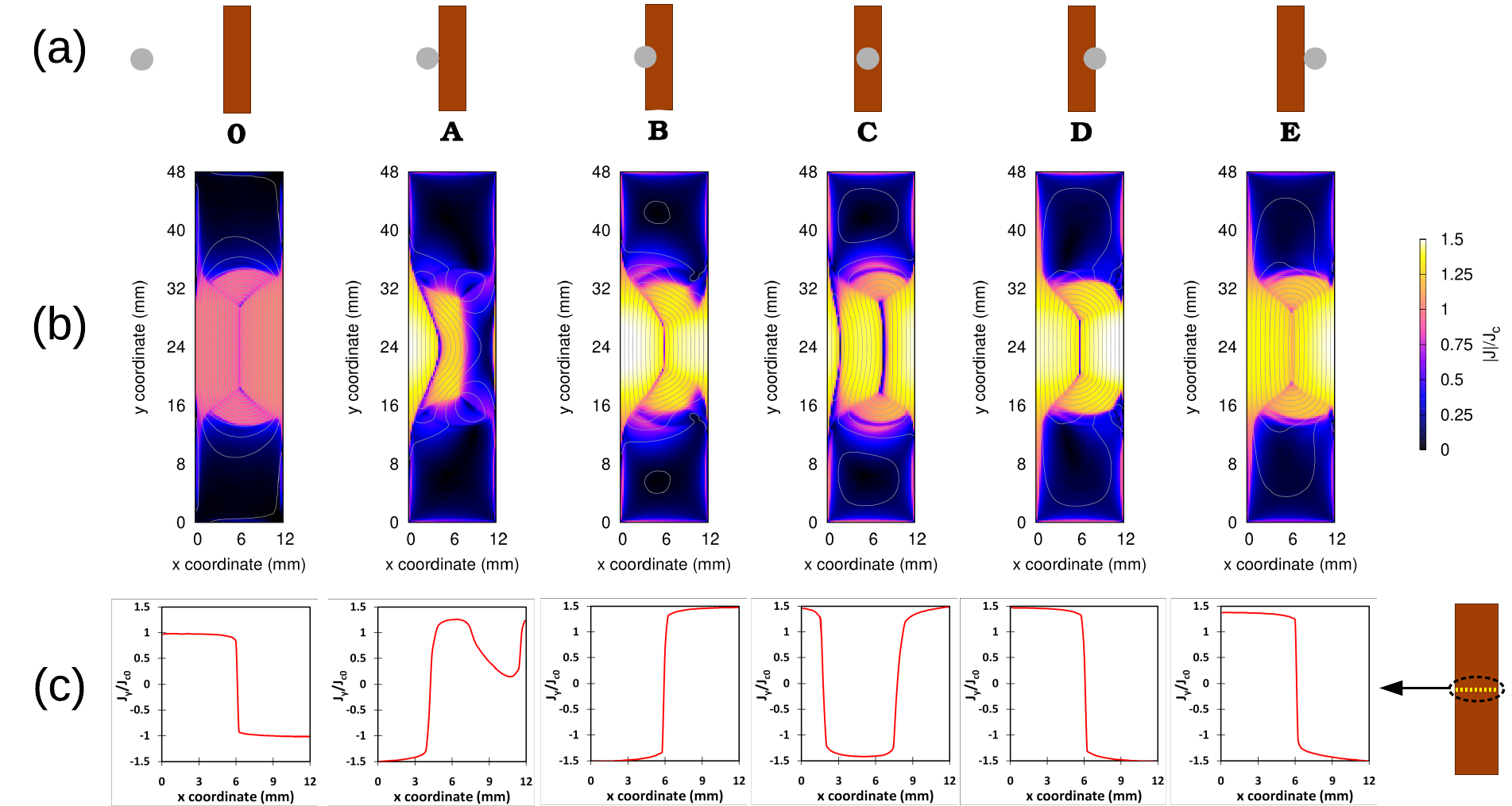}}
\caption {Current modeling results for airgap 3.3 mm and constant $J_c$. (a) Key magnet positions including step \textbf{0}, when magnet is very far from the tape. (b) Current modulus maps and current lines regarding the key magnet positions. (c) Current profiles of $J_c$-normalized $J_y$ regarding the key magnet positions in the mid-plane of the tape ($y=$24 mm).} 
\label{fig.J_steps}
\end{figure}

Fig. \ref{fig.J_steps}(b) displays the current density (maps of the modulus and current lines) for key magnet positions. Step \textbf{0} belongs to the beginning of second cycle when the magnet is still far away from the tape and the tape is fully saturated with screening currents remained from the first cycle. In this step, no overcritical currents in the tape are observed (see also Fig.\ref{fig.J_steps}(c) related to $J_c$-normalized $J_y$ profiles in the middle of the tape length). At step \textbf{A}, the magnet touches the tape and starts to enter it from the left side. Since the magnet already creates high applied magnetic fields on the tape, overcritical currents can be observed in the left side of the tape. These circulating currents are induced in such a way that their resultant magnetic field can oppose the magnetic field of magnet and eventually repel it (based on Lenz's law). At step \textbf{B}, half of the magnet has entered the tape and the magnet magnetic field  has occupied the whole tape. Thus, the overcritical currents can be observed throughout the tape. At step \textbf{C}, the magnet has reached the middle of the tape while the tape is still fully occupied with overcritical screening currents. The screening currents direction has been altered compared to step \textbf{B} because of changing the direction of the magnet magnetic field inside the tape. At step \textbf{D}, the situation is similar to step \textbf{B} but with opposite direction of overcritical screening currents. At step \textbf{E}, while the magnet is leaving the tape, the tape is still fully saturated with overcritical screening currents and the direction of these currents are the same as step \textbf{0}. While the magnet moves far enough from the magnet, the overcritical currents in the tape vanish completely and the remanent screening currents in the tape become similar to step \textbf{0}.

\begin{figure}[tbp]
\centering
{\includegraphics[trim=0 0 0 0,clip,width=12.5 cm]{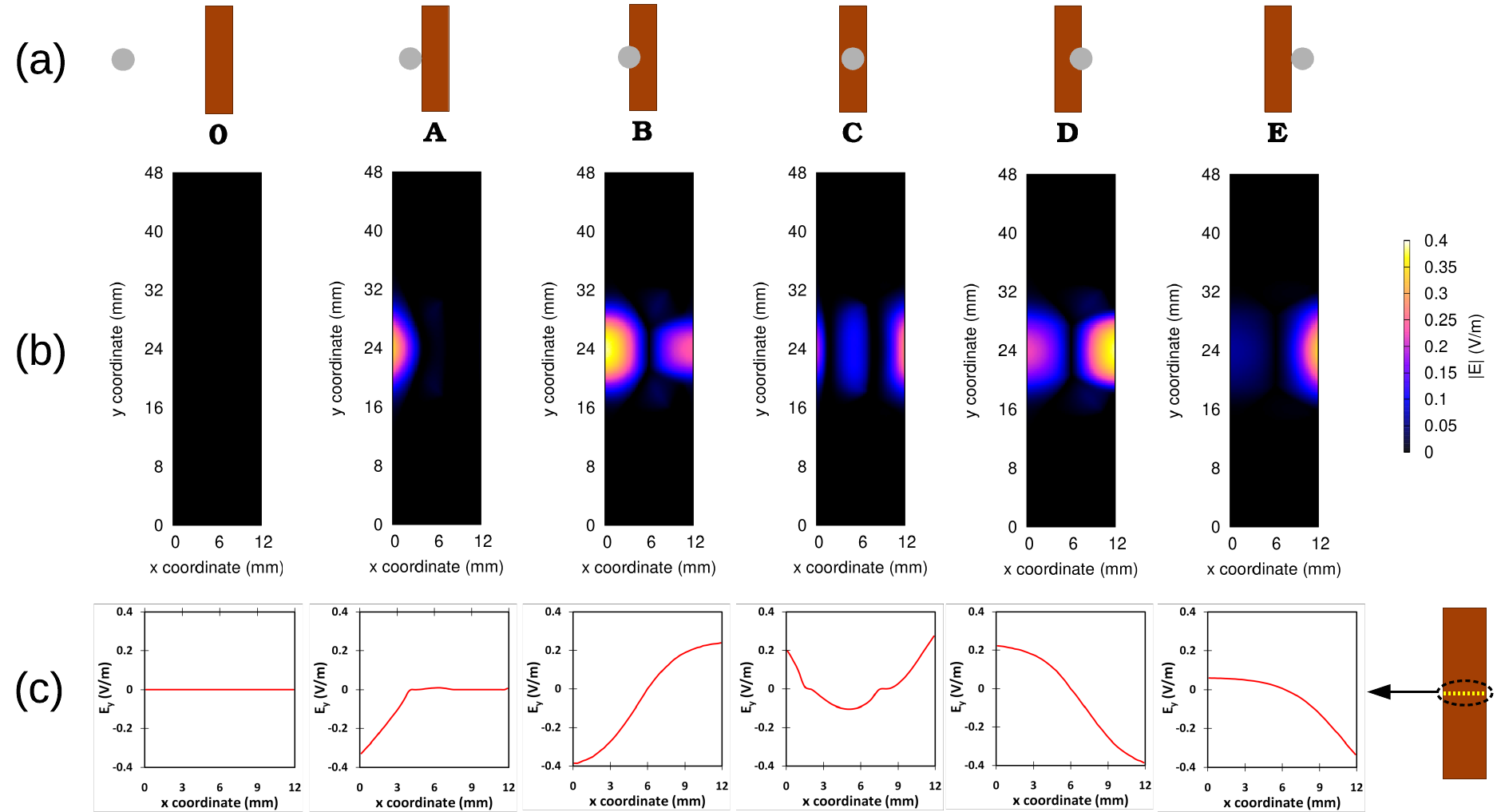}}
\caption {Electric Field modeling results for airgap 3.3 mm and constant $J_c$. (a) Key magnet positions. (b) Electric field modulus maps regarding the key magnet positions. (c) Electric field profiles of $E_y$ regarding the key magnet positions in the mid-plane of the tape ($y=$24 mm)..} 
\label{fig.E_steps}
\end{figure}

Far away from the magnet and the main screening currents, currents of lower magnitude (just below $J_c$) appear on the tape edges on the sides and, with even lower magnitude, at the ends. These currents have higher values along the left edge of the tape, where the magnet enters the tape and become maximum at step \textbf{D}. As it will be shown in Fig. \ref{fig.E_steps}, these low values of screening current have insignificant contribution in generating electric field in the tape. For example at step \textbf{D}, the electric field due to these low values of screening currents are 3 orders of magnitudes smaller than the maximum electric field in the tape at their highest values.                

Fig. \ref{fig.E_steps}(b) depicts electric field maps related to the key magnet positions and Fig. \ref{fig.E_steps}(c) shows the profiles of the $y$ component of the electric field, $E_y$, in the middle of the tape length (the $E_x$ component vanishes there). The main area in the tape responsible to generate voltage in the tape lies under and along the magnet cross section (area from 19 mm to 29 mm along $y$ axis), where the magnet is traversing. At steps \textbf{B} and \textbf{D} the largest yellow areas are observed, showing the high values of electric field induced in these areas (around 0.4 V/m), which justifies the occurrence of minimum peaks in Fig. \ref{fig.DeltaV_key} at these steps.          

\begin{figure}[tbp]
\centering
{\includegraphics[trim=0 0 0 0,clip,width=12.5 cm]{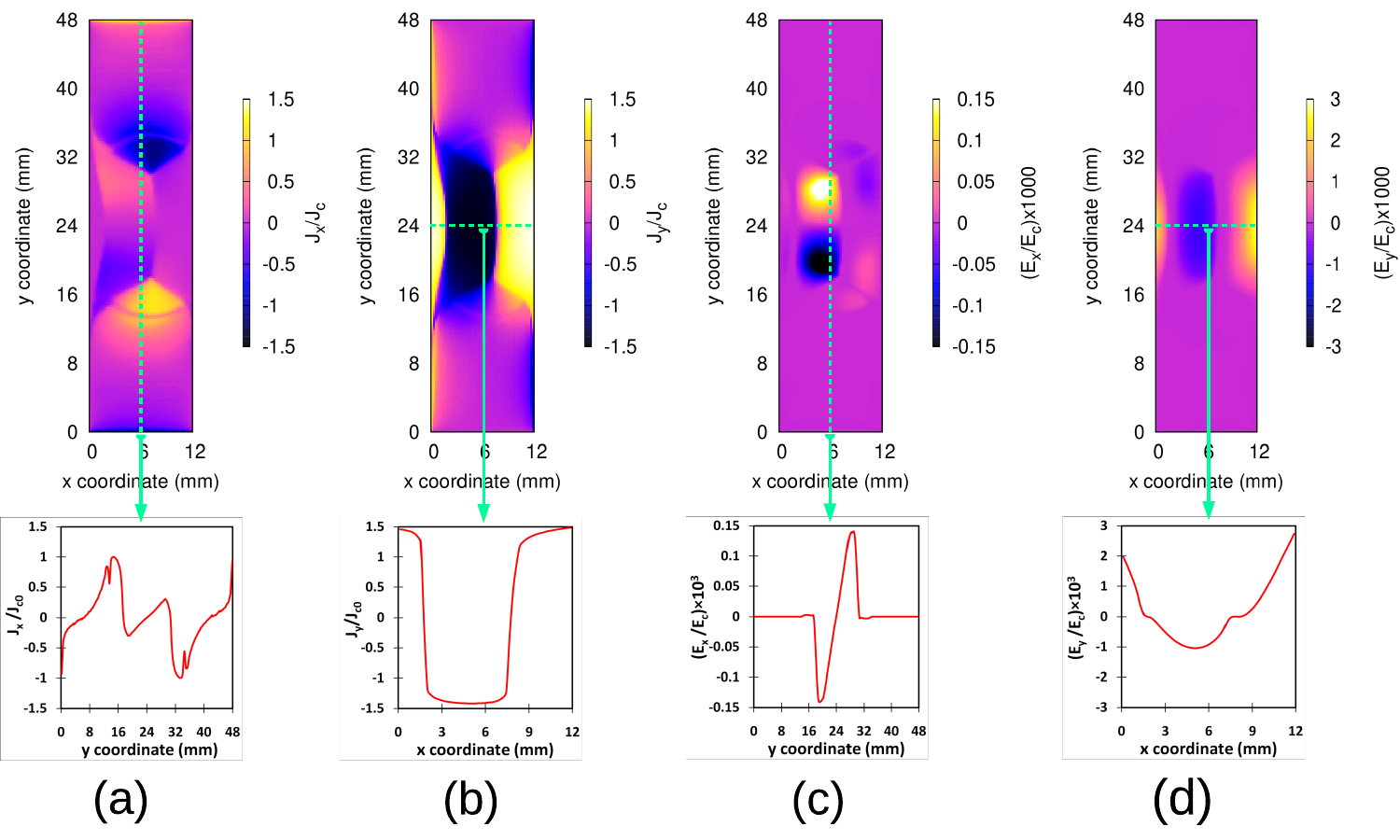}}
\caption {Modeling results for the $x$ and $y$ components of current density and electric field while the magnet is just on the top and concentric to the tape (step \textbf{C}) for airgap 3.3 mm and constant $J_c$. (a) $J_c$-normalized current map and current profile of $J_x$ along the $y$ axis in the middle of the tape (dashed green line). (b) $J_c$-normalized current map and current profile of $J_y$ along the $x$ axis in the middle of the tape (dashed green line). (c) $E_c$-normalized electric field map and electric field profile of $E_x$ along the $y$ axis in the middle of the tape (dashed green line). (d) $E_c$-normalized electric field map and electric field profile of $E_y$ along the $x$ axis in the middle of the tape (dashed green line).} 
\label{fig.EJ_xy}
\end{figure}

In Fig. \ref{fig.EJ_xy}, we study the $x$ and $y$ components of the current density and electric field in more detail in a certain time step, being when the magnet is just on the top and concentric to the tape (step \textbf{C}). In Fig. \ref{fig.EJ_xy}(a) it is obvious that $J_x$ stays just below the critical current while $J_y$ reaches up to 1.5 times of $J_c$. As a results of these, $E_y$ is almost 20 times higher than $E_x$ [see Fig.\ref{fig.EJ_xy}(c) and (d)]. In addition, since $J_x$ is symmetric along the length, $E_x$ is also symmetric [see Figs. \ref{fig.EJ_xy}(a) and (c)]. Thanks to this symmetry, the average $E_x$ in the whole sample vanishes, and thence $E_x$ does not contribute to the voltage.

In addition, it is worth mentioning that the the maximum value of $J_x$ occurs at steps \textbf{D} and \textbf{E} with maximum $J_c$-normalized values of 1.23 and 1.20, respectively. The resultant $E_x$ due to these currents is still insignificant compared to $E_y$ generated by $J_y$. This $E_x$ can be recognized as vague blue areas in Fig.\ref{fig.E_steps}(b) at steps \textbf{B} and \textbf{D} outside the magnet cross section area (above and below the area between 19 mm and 29 mm, where the magnet traverses).  

\subsection{Impact of Tape Length on Voltage Generation}

Studying the impact of the tape length on voltage generation assists us to realize what is the minimum efficient tape length (along the $y$ axis) that can be employed in a flux pump. From another point of view, this study help us to select the proper distance between voltage tabs while measuring the voltage signals in a flux pump. This distance is of course a variable of magnet dimension, especially along the length. In this section, we studied this effect with the same flux pump configurations for various airgaps. Now, we use $J_c(B,\theta)$ dependency in order to obtain a realistic description that is closer to experiments. The tape length, changing between 5 mm to 48 mm, and the airgap, varying from 1 mm to 10 mm, are chosen as variables while the frequency is kept constant as 12.3 Hz. 

\begin{figure}[tbp!]
\centering
{\includegraphics[trim=0 0 0 0,clip,width=10 cm]{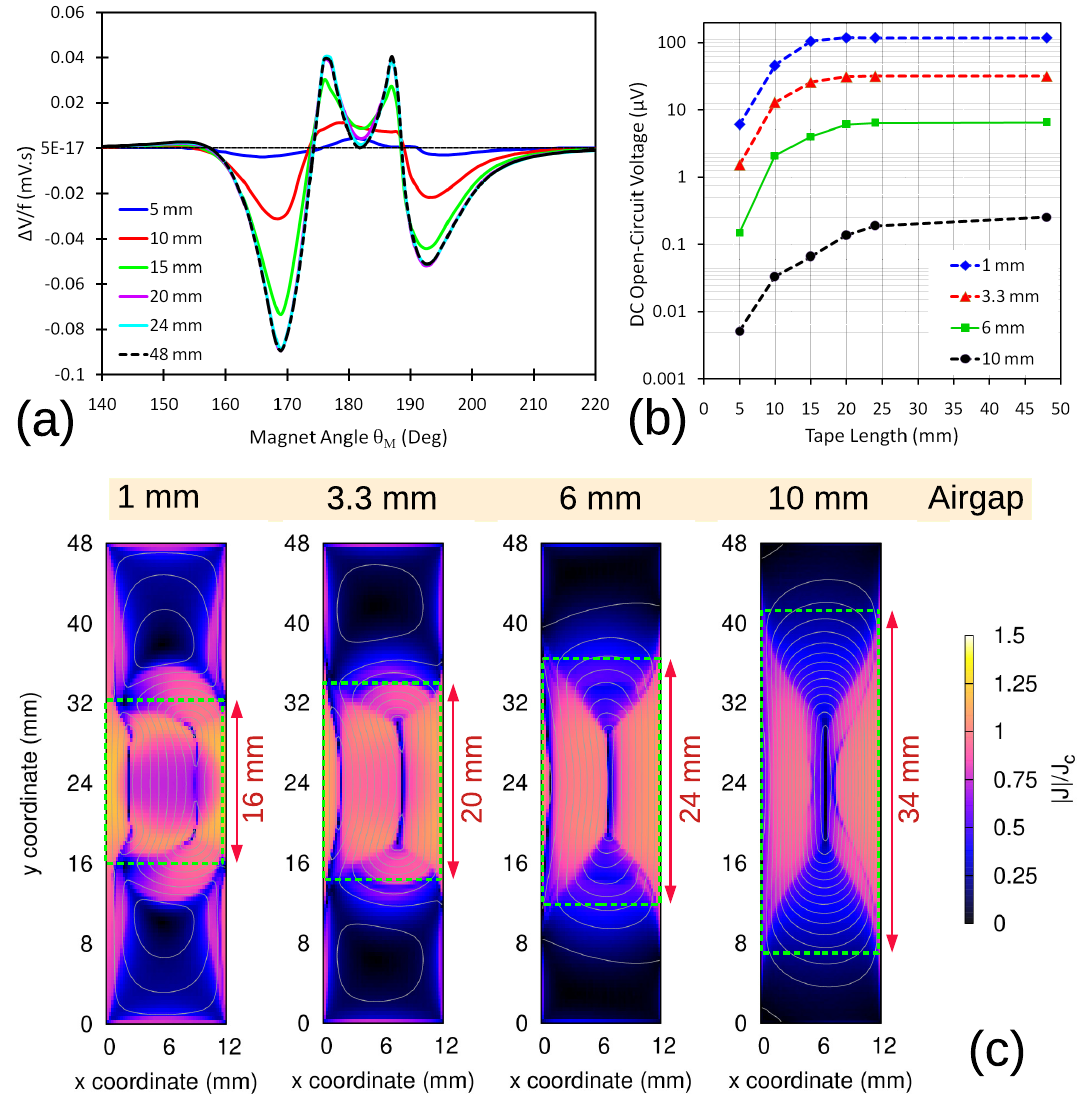}}
\caption {(a) Frequency-normalized $\Delta V$ against magnet angle $\theta_M$ for different tape lengths for airgap of 3.3 mm. (b) The DC open-circuit voltage value versus tape lengths for various airgaps. (c) Current modulus maps and current lines for different airgaps when the magnet is just in the middle and concentric to the tape ($\theta_M=180^{\circ}$).} 
\label{fig.Tape_length}
\end{figure}

Fig. \ref{fig.Tape_length}(a) shows the result of $\Delta V$ normalized by the frequency against the magnet angle $\theta_M$ for different tape lengths and for 3.3 mm airgap. It is clear that above 24 mm length, the results of $\Delta V$ remain almost the same and the curves coincide to each other. Also, Fig. \ref{fig.Tape_length}(b) shows the DC open-circuit voltage value versus tape length for various airgaps, from 1 mm to 10 mm. Up to the airgap of 6 mm, the DC voltage changes only a few percent with increasing the tape length from 24 to 48 mm. This can be justified by the distribution of current modulus on the tape surface for different airgaps, shown in Fig. \ref{fig.Tape_length}(c). In this figure, the dashed green rectangulars show the area on the tape surface containing the flow of $y$ component of current density, $J_y$, that is responsible for creation of voltage in the tape. Almost all of the created voltage in the tape is generated in this area. As the airgap increases, the rectangular becomes longer accordingly. This means that current distributes over a broader area across the tape, and thus the voltage is generated over a wider area. This fact also can be recognized in Fig. \ref{fig.Tape_length}(b) by the decrease in the slope of the curves from 1 mm to 10 mm airgap, showing that in smaller airgaps, the induced voltage is more concentrated in the center of the tape, where the magnet traverses.
       
For voltage signal measurements, the voltage tabs should be placed in any area of the tape outside the two sides of the rectangle in order to measure the voltage signal accurately. 

As a conclusion, up to airgap of 6 mm, the voltage tabs distance should be at least 2.5 times larger than the magnet length. However, higher than this airgap, this distance is not sufficient anymore and for example at the airgap of 10 mm, the voltage tabs distance should be around 3.5 times larger than the magnet length to be able to capture the voltage signal with highest accuracy. 
           
\subsection{Comparison to Experiments}
\label{s.comp_exp}

In this section, we compare our results of the modeled dynamo-type HTS flux pump with experiments published in \cite{bumby2016anomalous} and with 2D MEMEP method presented in \cite{Ghabeli_2020}. The calculations have been conducted with $J_c(B,\theta)$ dependency and with frequency of 12.3 Hz to be comparable with experimental results.

\begin{figure}[tbp!]
\centering
{\includegraphics[trim=0 0 0 0,clip,width=8 cm]{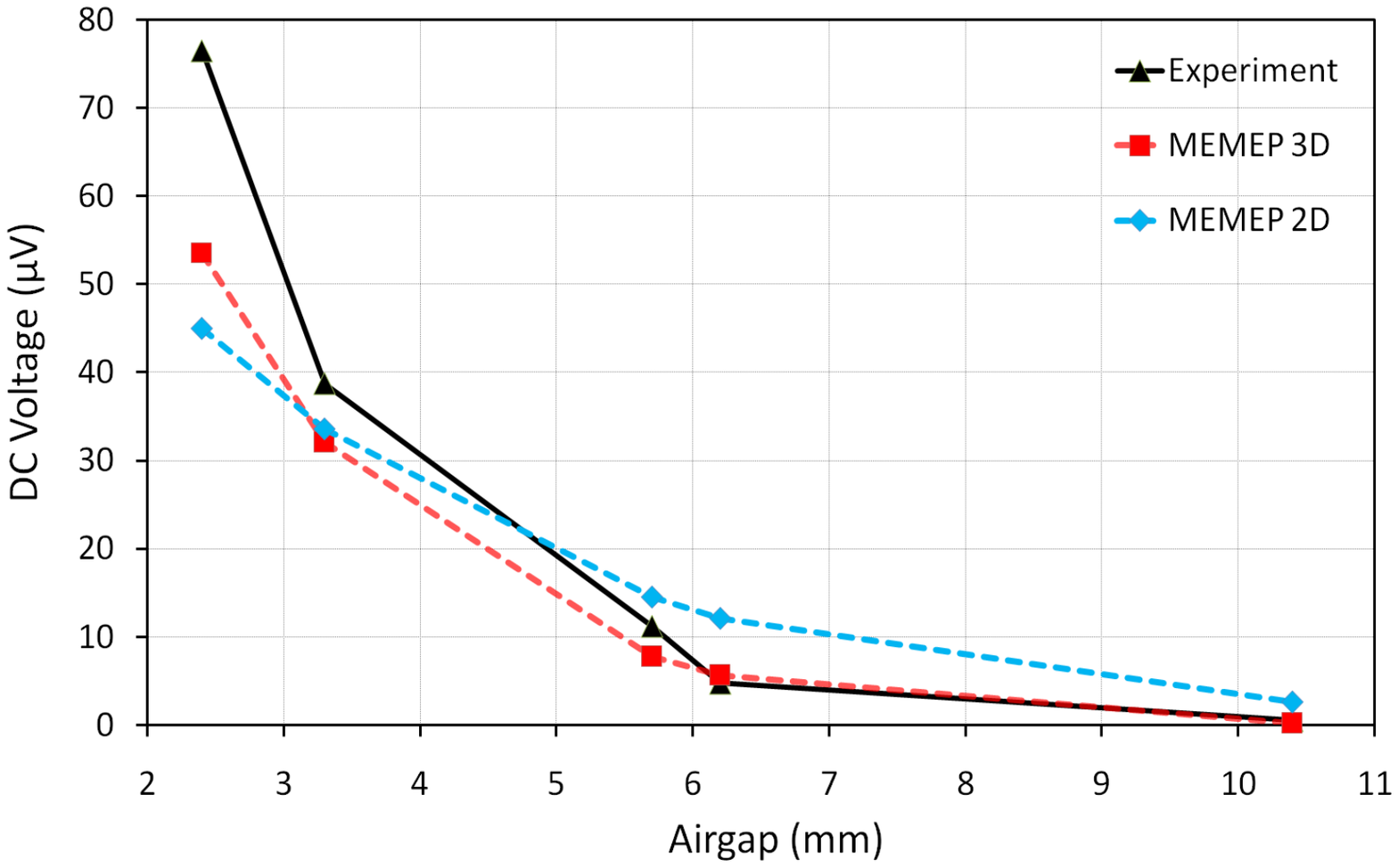}}
\caption {Comparison between DC open-circuit voltage values of different airgaps for calculated results of MEMEP 3D modeling against MEMEP 2D modeling presented in \cite{Ghabeli_2020} and experiments conducted in \cite{bumby2016anomalous}. Better agreement for 3D modeling is evident at large airgaps, while for low airgaps we expect significant measurement errors.} 
\label{fig.Airgap}
\end{figure}

 Fig. \ref{fig.Airgap} shows the results of the calculated and measured DC open-circuit voltage values for 5 different airgaps from 2.4 mm to 10.4 mm. Looking at the figure, it is obvious that the MEMEP 3D presents very good agreement with experiments and even better agreement than the MEMEP 2D method. In the case of MEMEP 2D, the decay of the DC voltage with the airgap is less pronounced than the measurements, with substantial differences at high airgaps. The reason is that the magnetic field of the magnet at 2D decays with the distance, $r$, as $1/r^2$ (infinitely long dipole), while for the real cylindrical magnet the magnetic field decays as $1/r^3$ (point dipole). This is the reason why 3D modeling agrees better with experiments. It is worth mentioning that, as discussed also in \cite{Ghabeli_2020}, the airgap measurement error is not negligible due to contraction and mechanical instability of the flux pump in liquid nitrogen bath (at least $\pm$0.5 mm airgap). This error is mostly noticeable in the lowest airgap of 2.4 mm, where a small difference in the airgap causes a high difference in the magnetic field from the magnet, and hence DC voltage. Should the gap be more accurately measured, we expect better agreement between model and experiments at low gaps. Moreover, another source of discrepancy can be that the $J_c(B,\theta)$ in the model does not exactly correspond to that of the measurements, since the original tape is no longer available. As well, we needed to assume that $J_c(B,\theta,\phi)=J_c(B,\theta,\pi/2)$ for any $\phi$ because of unavailable measurements of this characteristics (see section \ref{HTS Tape modeling}).

\section{Conclusion}

In this article, the first 3D model of an HTS flux pump has been presented and verified against experiment. This model can help to clarify the complicated mechanism of HTS dynamo-type flux pumps and verify the previously proposed mechanism by 2D models. The unique features of the presented 3D model, based on Minimum Electro-Magnetic Entropy Production (MEMEP), makes the model very fast and efficient (solving the Maxwell equations only inside the superconductor, fast solver by division into sectors and iteration, and use of parallel computing). Employing this model, we studied the screening current distribution, which in its overcritical form is the reason for flux pumping phenomenon, across the tape surface and in several key positions of magnet movement. We investigate the role of the $x$ and $y$ components of electric field and screening current in generating the voltage in the flux pump. It was found that, while the magnet traverses the tape, the maximum value of the $y$ component of the electric field, $E_y$, is around 20 times larger than $E_x$, which highlights the role of $J_y$ to generate the voltage throughout the tape surface. In addition, the average $E_x$ across the sample vanishes because of symmetry. We also studied the effect of tape length on voltage generation in the flux pump. We found that the distance between voltage tabs for precise measuring of the voltage depends on not only the diameter of the magnet but also the distance between magnet and tape surface. This distance should be at least 2.5 times and 3.5 times larger than the magnet diameter for airgaps less than 0.6 and around 1 times the diameter, respectively. We expect that the minimum tape length will keep increasing with the separation. Finally, the calculated modeling results were compared against experiments for several airgaps, which showed very good agreement.           
  
\section*{Acknowledgments}
This work received the financial support of the Slovak grant agencies APVV with contract number APVV-19-0536 and VEGA with contract number 2/0097/18.

\section*{Author contributions statement}

AG performed the modeling and developed the specific routines for flux pumping; EP conceived the simulations background, provided physical insight, and leaded the research; MK and EP developed a previous version of the 3D modeling software; AG and EP wrote the article; AG prepared all figures.

\section*{Additional information}

\textbf{Competing interests:} The authors declare that they have no competing interests. 



\end{document}